          [2001/04/25 1.1 (PWD)]
\documentclass[a4paper,twocolumn]{esapub} 

\usepackage{natbib}
\usepackage{graphicx}
\usepackage{url}

\title{The INTEGRAL/SPI response and the Crab observations}
\author[1]{P.~Sizun}
\affil[1]{CEA Saclay, DSM/DAPNIA/Service d'Astrophysique, 91191
Gif-sur-Yvette, France}
\author[2]{C.~R.~Shrader}
\affil[2]{NASA/Goddard Space Flight Center, MD 20771, Greenbelt,
USA}
\author[1]{D. Atti\'{e}}
\author[3]{P. Dubath}
\affil[3]{INTEGRAL Science Data Center, Chemin d'Ecogia 16, 1290
Versoix, Switzerland}
\author[1]{S.~Schanne}
\author[1]{B.~Cordier}
\author[2]{S.~J. Sturner}
\author[4]{L.~Bouchet}
\author[4]{J.-P.~Roques}
\author[4]{G.~K.~Skinner}
\affil[4]{Centre d'Etude Spatiale des Rayonnements, 31028
Toulouse, France}
\author[5]{P.~Connell}
\affil[5]{GACE - ICUMV, Universitat de Val\`{e}ncia, Apdo 22085,
46071 Val\`{e}ncia, Spain}

\begin{document}


\maketitle

\begin{abstract}
The Crab region was observed several times by INTEGRAL for
calibration purposes. This paper aims at underlining the
systematic interactions between (i) observations of this reference
source, (ii) in-flight calibration of the instrumental response
and (iii) the development and validation of the analysis tools of
the SPI spectrometer \citep{vedrenne2003}. It first describes the
way the response is produced and how studies of the Crab spectrum
lead to improvements and corrections in the initial response.
Then, we present the tools which were developed to extract spectra
from the SPI observation data and finally a Crab spectrum obtained
with one of these methods, to show the agreement with previous
experiments. We conclude with the work still ahead to understand
residual uncertainties in the response.
\end{abstract}

\section{Crab observations}
During its first year in orbit, INTEGRAL observed the Crab region
(the nebula and its pulsar) twice : in February (revolutions
39-45) and August (revolutions 102-103) 2003. The total
observation time yields 1,723 ks (including annealing periods),
performed in different modes : on- or off-axis, in staring or
dithered mode, with a 5 x 5 or an hexagonal dither pattern.
Because of its high luminosity and apparent lack of variability,
this source is used as a reference for the spectrometer
calibration.

\section{Instrumental response}

\paragraph{Response decomposition}
The response is decomposed into IRFs and RMFs :
\begin{itemize}
\item[-] the redistribution matrix files (RMFs) are divided into
three components describing (1) the events of the full-energy
peak, (2) the Compton events interacting first in the detectors,
(3) those interacting first in the passive material. They do not
depend on the direction or the detector but only on the energy.

\item[-] the image response files (IRFs) describe, for a given input
photon energy, the effective area of each detector for all
directions in the field of view.
\end{itemize}

\paragraph{Production}
The INTEGRAL/SPI instrument response is produced using a suite of
Monte Carlo simulation software developed at NASA/GSFC based on
the GEANT-3 package called MGEANT \citep{sturner2003}. This
production also required the development of a detailed computer
mass model for SPI.

\begin{figure}[!bhtp]
\includegraphics[width=0.9\linewidth]{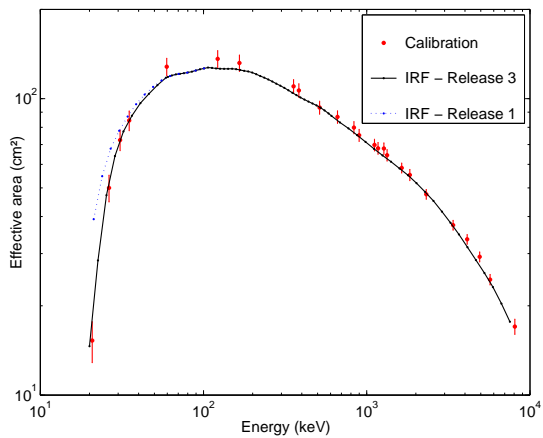}
\caption{\label{fig:irfs}SPI photopeak effective area : comparison between ground
calibration measurements (dots with error bars), and two
successive IRF releases (dotted and plain lines).}
\end{figure}

\newpage
Absolute determination of the camera photo-peak effective area was
obtained from measurements (points in Fig.~\ref{fig:irfs})
performed just before the launch during an extensive ground
calibration campaign \citep{attie2003,schanne2003} of SPI in 2001
April at the Bruy\`{e}res-le-Ch\^{a}tel (BLC) site of CEA. These
measurements were compared to simulations and led to the initial
release of the SPI response (Fig.~\ref{fig:irfs}, dotted line).

\paragraph{Correction}
Initial analysis of ground calibration data was performed only on
lines with energies at or above 60~keV. After the launch, it was
found through analyses of early Crab observations, that the Crab
flux below these 60~keV was underestimated compared to the
spectrum expected. It enabled us to point out an over-estimation
of the low energy efficiency.

Additional analyses of 4 low-energy lines from the BLC data were
performed. The $20.80$ and $26.35$~keV lines of the $^{241}$Am
calibration source had initially been ignored because of their low
statistics. The $\sim30.8$ and $\sim35.07$ keV lines of
$^{133}$Ba, blended by the Compton component of higher energy
lines, had also been rejected. A new response, corrected below 60
keV by taking into account these 4 lines, was derived
(Fig.~\ref{fig:irfs}, plain line), without any assumption on the
Crab spectrum.

\section{Deconvolution methods}
\begin{figure}[!hbtp]
\includegraphics[width=0.9\linewidth]{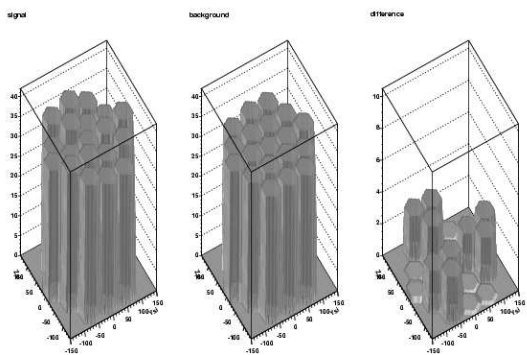}
\caption{\label{fig:3d}Count rates in the 19 Ge detectors in the 20-8000 keV
band, during a Crab on axis pointing. From left to right : total
count rate, estimated background count rate and their difference.
The pattern of the projection of an on axis source through the
mask on the detector plane, with its 120 degrees symmetry, is
visible.}
\end{figure}

Several tools have been developed to extract source spectra from
SPI observation data, including Spiros and XSPEC 12.
\begin{itemize}
\item[-] in its spectral mode, Spiros \citep{skinner2003},
available in the INTEGRAL off-line scientific analysis (OSA)
distribution (\url{http://isdc.unige.ch/index.cgi?Soft+soft}),
adjusts the intensity of sources simultaneously with the scaling
factors to apply to the background model chosen. The best solution
is searched for in each energy bin successively. The off-diagonal
terms of the instrumental response are not taken into account in
the deconvolution -- all events are treated as photopeak events --
and the resulting spectrum is in \emph{pseudo-photons}.
 A ``\emph{Spiros}
dedicated'' redistribution matrix was derived using Monte Carlo
simulations of both the SPI instrument and the \emph{Spiros}
software. This matrix can for example be used by the
spectral-fitting program \emph{XSPEC}, to fit a model to the real
photon spectrum ;
\item[-] \emph{XSPEC~12}, developped at NASA/GSFC \citep{shrader2004} is a new release
of \emph{XSPEC} including SPI specific packages -- to be delivered
soon. While \emph{XSPEC 11} can be used to fit a mathematical
model to an already background substracted and deconvolved SPI
source spectrum, \emph{XSPEC 12} can work directly from the
individual detector spectra. Using, unlike \emph{Spiros}, the full
intrumental response -- IRFs and RMFs -- reconstructed for each
source, it adjusts simultaneously the model parameters of each of
the sources considered and the scaling factors to apply to the
background model.
\end{itemize}

To account for initial discrepancies in the Crab derived flux
between \emph{Spiros} and \emph{XSPEC 12}, a new version of the
``\emph{Spiros} dedicated'' redistribution matrix mentioned above
was recently derived at NASA/GSFC.

\begin{figure}[!hbtp]
\includegraphics[width=0.9\linewidth]{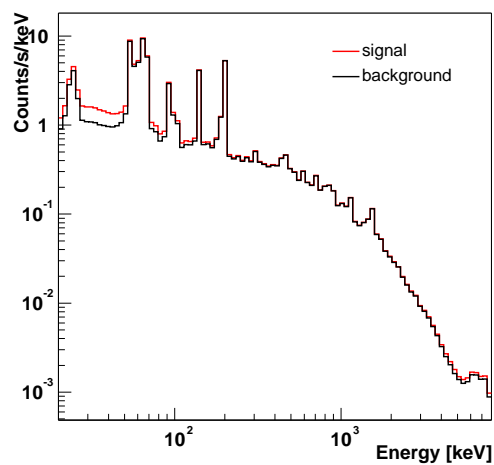}
\caption{\label{fig:bkg}Total count rate (upper line) and estimated
background count rate (lower line) during a Crab observation : the
signal of interest represents only a tiny fraction of the total
count rate.}
\end{figure}

However, the main challenge is to properly model the background
which represents a huge fraction of the total measured count rate
(Fig.~\ref{fig:bkg} and 3).

\begin{center}
\begin{figure*}[!thbp]
\begin{center}
\includegraphics[angle=270,width=0.8\linewidth]{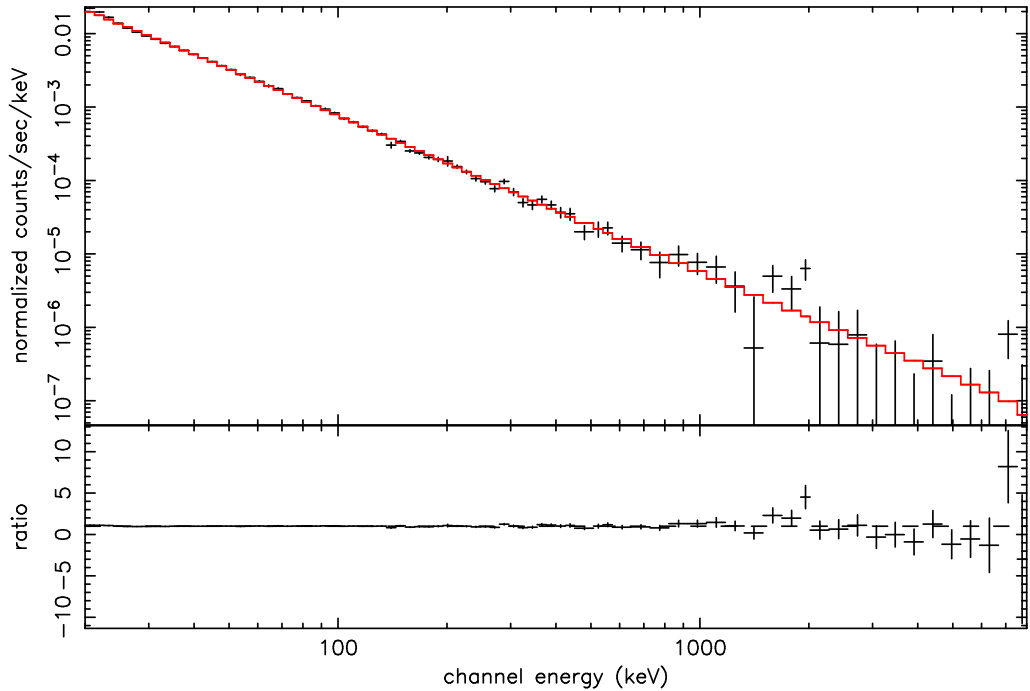}
\caption{\label{fig:spec}SPI Crab spectrum extracted from 44 NRT science windows in
INTEGRAL revolution 44, using OSA 3.0 and a saturated Ge
background model fitted detector by detector with Gaussian
statistics.}
\end{center}
\end{figure*}
\end{center}

\newpage
\noindent Spectral studies of Crab and other reference sources
were done with both \emph{Spiros} and \emph{XSPEC 12}, enabling us
to cross-calibrate these tools. The resulting spectra are now
consistent in shape and normalization. An initial discrepancy in
normalization was reduced after a new redistribution matrix was
derived for \emph{Spiros}. Some differences remain which are still
under study.

\section{Crab spectrum}
The Crab spectrum in Fig.~\ref{fig:spec} was extracted using as
standard a method as possible, so that any scientist undertaking
to analyze SPI observations shall be able to reproduce such a
spectrum. It corresponds to the first 44 near real time science
windows of revolution 44, performed with a 5 $\times$ 5 dither
pattern. Distribution 3.0 of the observation analysis software
(OSA) was used and the \emph{spi\_science\_analysis} script was
run mainly with default parameters, the most important of which
concern the background~: spiback produced a background model based
on the count rates in saturating Ge detectors and \emph{Spiros}
adjusted the source flux and background count rates in each energy
bin, assuming the time variation given by the model and computing
the best detector to detector ratios.

\begin{center}
    \begin{table}[!bhtp]
        \begin{center}
        \caption{\label{tab:flux}Flux (ph/cm$^2$/s) and power law index of total
Crab gamma-ray emission.}
        \begin{tabular}{l|c|c}
        \hline
         Instrument &   Index & 50-100 keV flux \\
         \hline
         OSO-8      &   $2.00 \pm 0.06$     & $6.41 \,10^{-2}$\\
         GRIS       &   $2.15 \pm 0.03$     & $4.52 \,10^{-2}$\\
         CGRO/OSSE$^\dag$ & $2.19 \pm 0.03$  & $5.68 \,10^{-2}$\\
         CGRO/BATSE$^\dag$& $2.20 \pm 0.01$  & $6.83 \,10^{-2}$\\
         SAX/PDS$^\dag$   & $2.13 \pm 0.01$  & $4.92 \,10^{-2}$\\
         INTEGRAL/SPI     & $2.17 \pm 0.01^\ast$ & $7.08 \pm 0.03^\ast \,10^{-2}$\\
         \hline
        \end{tabular}
        \end{center}
        $^\dag$ private communication \\
        $^\ast$ statistical error only
    \end{table}
\end{center}

\newpage
\noindent Above 1~MeV, the flux uncertainty grows up as the source
flux gets smaller and the basic background model used starts to
show its limits~: the features visible above 1~MeV correspond to
instrumental background residuals.

\newpage
\noindent A power law $F(E) = K
\left(\frac{E}{1~\mathtt{keV}}\right)^{-\alpha}$ with a spectral
index $\alpha = 2.169 \pm 0.008$ and a normalization $K = 14.44
\pm 0.44$ ph/cm$^2$/s/keV fits well to the spectrum
($\chi^2$/d.o.f $= 1.31$) in the 40~keV to 8~MeV energy range.

\noindent Imposing the canonical photon spectral index of $2.10$
found by X-ray experiments for the entire Crab, we estimate
($\chi^2$/d.o.f $= 2.39$) a normalization of $10.80 \pm 0.03$
ph/cm$^2$/s/keV at 1~keV, to compare with a value of 9.59
ph/cm$^2$/s/keV given by \citet{willingale2001}.

\noindent A broken power law fits only slightly better
($\chi^2$/d.o.f $= 1.29$) than a single power law. Imposing a low
energy index $\alpha_1 = 2.10$, we find a high energy index
$\alpha_2 = 2.19 \pm 0.01$ with a break around $61 \pm 6$~keV and
a normalization of $11.03 \pm 0.05$ ph/cm$^2$/s/keV at 1~keV.

\noindent Using a different energy range or background handling
method, the fit parameters found vary slightly.
 Even without taking into account systematic uncertainties due to the
calibration and background handling works still in progress,
Table~\ref{tab:flux} shows a very good agreement between SPI and
other gamma-ray experiments.

\section{Conclusion}
After a few months of fine tuning of both the instrumental
response and the deconvolution tools, the Crab spectra extracted
from INTEGRAL/SPI observations with the various software available
are compatible with each other and rather consistent with previous
experiments.

\noindent Some further work will be necessary to build more
elaborate background models and become more confident in the
fluxes derived. Although the response and the extraction software
might still evolve in the future – especially to account for the
loss of detector 2, the tools available today are sufficient to
study point sources observed with INTEGRAL.

\section*{ACKNOWLEDGEMENTS}
We thank the team responsible for the SPI spectrometer aboard
INTEGRAL for their contribution to this work. INTEGRAL is an ESA
project with instruments and science data centre funded by ESA
member states (especially the PI countries~: Denmark, France,
Germany, Italy, Switzerland, Spain), Czech Republic and Poland,
and with the participation of Russia and the USA.

\bibliographystyle{aa}
\bibliography{Sizun}

\begin{thebibliography}{10}
\expandafter\ifx\csname natexlab\endcsname\relax\def\natexlab#1{#1}\fi
\expandafter\ifx\csname url\endcsname\relax
  \def\url#1{{\tt #1}}\fi
\expandafter\ifx\csname urlprefix\endcsname\relax\def\urlprefix{URL }\fi

\bibitem[{{Atti{\' e}} et~al.(2003){Atti{\' e}}, {Cordier}, {Gros}
  et~al.}]{attie2003}
{Atti{\' e}} D., {Cordier} B., {Gros} M., et~al., Nov. 2003, A\&A, 411, L71

\bibitem[{{Bartlett} et~al.(1993){Bartlett}, {Barthelmy}, {Gehrels}
  et~al.}]{bartlett1993}
{Bartlett} L.M., {Barthelmy} S.J., {Gehrels} N., et~al., 1993, In: AIP Conf.
  Proc. 304, 67--71

\bibitem[{{Dolan} et~al.(1977){Dolan}, {Crannell}, {Dennis} et~al.}]{dolan1977}
{Dolan} J.F., {Crannell} C.J., {Dennis} B.R., et~al., Nov. 1977, ApJ, 217, 809

\bibitem[{MGEANT()}]{mgeant}
MGEANT, A generic multi-purpose monte-carlo package for gamma-ray experiments,
  {URL}: {\url{http://lheawww.gsfc.nasa.gov/docs/
  gamcosray/legr/mgeant/mgeant.html}}

\bibitem[{{Schanne} et~al.(2003){Schanne}, {Cordier}, {Gros}
  et~al.}]{schanne2003}
{Schanne} S., {Cordier} B., {Gros} M., et~al., Mar. 2003, In: Proc. SPIE, 4851,
  1132--1143

\bibitem[{{Shrader}(2004)}]{shrader2004}
{Shrader} C.R., 2004, In: this volume

\bibitem[{{Skinner} \& {Connell}(2003)}]{skinner2003}
{Skinner} G., {Connell} P., Nov. 2003, A\&A, 411, L123

\bibitem[{{Sturner} et~al.(2003){Sturner}, {Shrader}, {Weidenspointner}
  et~al.}]{sturner2003}
{Sturner} S.J., {Shrader} C.R., {Weidenspointner} G., et~al., Nov. 2003, A\&A,
  411, L81

\bibitem[{{Vedrenne} et~al.(2003){Vedrenne}, {Roques}, {Sch{\" o}nfelder}
  et~al.}]{vedrenne2003}
{Vedrenne} G., {Roques} J.P., {Sch{\" o}nfelder} V., et~al., Nov. 2003, A\&A,
  411, L63

\bibitem[{{Willingale} et~al.(2001){Willingale}, {Aschenbach}, {Griffiths}
  et~al.}]{willingale2001}
{Willingale} R., {Aschenbach} B., {Griffiths} R.G., et~al., Jan. 2001, A\&A,
  365, L212

\end{thebibliography}
\nocite{*}

\end{document}